\documentclass[%
 reprint,
superscriptaddress,
 amsmath,amssymb,
 prb,
]{revtex4-1}

\usepackage{dsfont}
\usepackage{graphicx}
\usepackage{hyperref}
\usepackage{color}

\usepackage{mathrsfs}
\usepackage{braket}

\usepackage{overpic}
\usepackage{tikz}
\usetikzlibrary{decorations.pathreplacing}
\usepackage[normalem]{ulem}

\newcommand{\im}{\mathrm{i}}

\definecolor{mygreen}{rgb}{0,0.5,0}
\definecolor{myblue}{rgb}{0,0,0.75}
\definecolor{mymagenta}{cmyk}{0,1,0,0.12}

\usepackage{umoline}

\begin{document}

\title{Tripartite information, scrambling, and the role of Hilbert space partitioning in quantum lattice models}

\author{Oskar Schnaack}
\author{Niklas B\"olter}
\author{Sebastian Paeckel}
\author{Salvatore R. Manmana}
\author{Stefan Kehrein}
\affiliation{%
 Institute for Theoretical Physics, 
	Georg-August-Universit\"at G\"ottingen, 
	Friedrich-Hund-Platz 1 - 37077 G\"ottingen, Germany
}
\author{Markus Schmitt}
\email{markus.schmitt@berkeley.edu}
\affiliation{%
 Department of Physics, University of California, Berkeley, CA 94720, USA
}
\affiliation{%
 Max-Planck-Institute for the Physics of Complex Systems, 
	N\"othnitzer Stra\ss e 38 - 01187 Dresden, Germany
}
\affiliation{%
 Institute for Theoretical Physics, 
	Georg-August-Universit\"at G\"ottingen, 
	Friedrich-Hund-Platz 1 - 37077 G\"ottingen, Germany
}
\date{\today}

\begin{abstract}
For the characterization of the dynamics in quantum many-body systems the question how information spreads and becomes distributed over the constituent degrees of freedom is of fundamental interest. The delocalization of information under many-body dynamics has been dubbed \emph{scrambling} and out-of-time-order correlators were proposed to probe this behavior. In this work we investigate the time-evolution of tripartite information as a natural operator-independent measure of scrambling, which quantifies to which extent the initially localized information can only be recovered by global measurements. Studying the dynamics of quantum lattice models with tunable integrability breaking we demonstrate that in contrast to quadratic models generic interacting systems scramble information irrespective of the chosen partitioning of the Hilbert space, which justifies the characterization as \emph{scrambler}.
Without interactions the dynamics of tripartite information in momentum space reveals unambiguously the absence of scrambling.
\end{abstract}

\maketitle

\section{Introduction.}
The concept of scrambling was originally devised to study the information paradox of black holes \cite{Hayden2007,Sekino2008}.
A scrambler is a quantum system with many degrees of freedom in which information about local fluctuations in the initial state is under dynamics strongly mixed up such that it can after long times only be recovered by global measurements.
It was found that black holes can be regarded as the most efficient scramblers \cite{Maldacena2016}.
The idea of scrambling is of interest also in quantum many-body systems beyond the AdS/CFT paradigm, where the spreading of correlations and information is a subject of ongoing research \cite{Lieb1972,Calabrese2005,Nachtergaele2010,Laeuchli2008,Manmana2009,Cheneau2012,Medvedyeva2013,Cevolani2016,Abeling2017} as well as the question of thermalization after a system was prepared far from equilibrium \cite{Gogolin2016,DAlessio2016} and how information about the initial conditions is lost \cite{Fine2014,Elsayed2015,Zangara2015,Schmitt2016,Schmitt2017,Schmitt2018,Hamazaki2018,Tarkhov2018,Yan2019,Pappalardi2019}.
Since the timescales of thermalization and scrambling can strongly differ, a central question is whether there is nevertheless a connection between both \cite{Bohrdt2017}.

In order to investigate scrambling from an information-theoretical point of view Hosur \emph{et al.} \cite{Hosur2016} introduced tripartite information as a measure for the delocalization of information. The tripartite information quantifies how much of the information about fluctuations that were in the initial condition localized in one part of the system can only be recovered when having access to both constituents of a bipartition of the time-evolved system. As such, tripartite information can be regarded as a direct probe of scrambling. A particular virtue is the fact that this information measure does not rely on any selection of operators. The only choice is the partitioning of the Hilbert space with respect to which it is decided whether information is distributed or not.
In this work, particular attention will be paid to the role of partitionings of Hilbert space in connection with the behavior of the tripartite information.
Note also Refs.\ \cite{Banuls2017,Iyoda2017,Zhou2017}, where alternative operator-independent measures for the spreading of information are investigated.

By contrast to information theoretic measures, so-called out-of-time-order correlators (OTOCs) of the form
\begin{align}
    C_{\hat V\hat W}=\braket{\hat V(t)^\dagger\hat W(0)^\dagger\hat V(t)\hat W(0)}_\beta\ ,
\end{align}
introduced in Refs. \cite{Shenker2014,Kitaev2014,Elsayed2013,Fine2014}, constitute an operator-based probe of scrambling.
In the expression above $\hat V(t)$ and $\hat W(t)$ are operators in the Heisenberg picture and $\braket{\cdot}_\beta$ denotes a thermal expectation value.
Considering local operators $\hat V_{A}$ and $\hat W_{B}$ acting on disjoint regions $A$ and $B$ the OTOC probes how the perturbation at $A$ affects the system at $B$ at later times.
In systems that scramble the perturbation eventually disturbs the whole system, which can be probed by the OTOC.
Moreover, considering the OTOC of momentum and position operator a semiclassical analysis motivates that OTOCs can indicate a butterfly effect in quantum systems, including the possible identification of Lyapunov exponents \cite{Larkin1969,Maldacena2016}.

With regard to the question of scrambling it is particularly notable that in spin-1/2 systems there exists a rigorous relation between 
OTOCs and tripartite information in the limit of high temperatures. In that case OTOCs bound the tripartite information such that the butterfly effect as diagnosed by an OTOC implies scrambling as measured by tripartite information \cite{Hosur2016}. This justifies to draw conclusions about the scrambling of information from the dynamics of OTOCs.

OTOCs have been studied in a series of works as a probe for the spreading of information and scrambling in condensed matter systems
\cite{Hosur2016,HuangY2016,Bohrdt2017,Iyoda2017,Swingle2016,Swingle2017,Tsuji2017,Fan2017,vonKeyserlingk2017,Rakovszky2017,Luitz2017,Nahum2017,Garttner2017,Lin2018,Xu2018,Khemani2018,Rammensee2018,Pappalardi2018,Sahu2018,Knap18}. 
However, tripartite information as a direct measure of the dispersion of information under dynamics has so far only been investigated in large-$N$ or long range interacting models \cite{Pappalardi2018,Seshadri2018,snderhauf2019quantum}.
%not received much attention
%\footnote{Tripartite information was used to investigate scrambling in Refs.\ \cite{Pappalardi2018,Seshadri2018,snderhauf2019quantum}.}.
The work presented in this paper comprises a systematic study of tripartite information as a measure of scrambling in quantum lattice models, with a particular focus on the role played by the choice of the partitioning of Hilbert space.
We demonstrate numerically that in the dynamics of generic interacting systems the tripartite information at late times approaches a stationary value that is close to the one obtained by evolution with a Haar random unitary. This behavior is independent of the chosen partitioning of the Hilbert space, indicating scrambling of information. By contrast, the time evolution of the tripartite information in quadratic systems varies with the Hilbert space partitioning and can be much smaller than the Haar value; hence, the dynamics of these systems cannot be regarded as scrambling.
The characteristic distinction between non-interacting and interacting systems in the view of scrambling is particularly pronounced in momentum space, for which to the best of our knowledge no results have been reported so far.

\section{Tripartite information.}
\begin{figure}[!t]
%\begin{overpic}[width=.5\columnwidth]{trip_inf.pdf}
%\put(44,45.5){$\hat U(t)$}
%\put(18,89){$A$}
%\put(59,89){$B$}
%\put(30,2){$C$}
%\put(70,2){$D$}
%\put(-5,20){out}
%\put(-4,73){in}
%\end{overpic}
\includegraphics[width=.6\columnwidth]{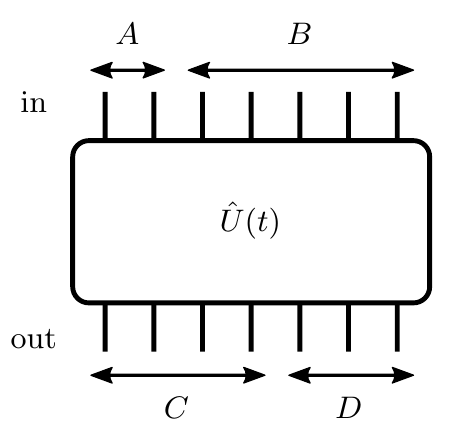}
%\begin{tikzpicture}
%\node at (0,0) {\includegraphics[width=.5\columnwidth]{trip_inf.pdf}};
%\node at (0,0) {$\hat U(t)$};
%\node at (-1.25,1.9) {$A$};
%\node at (.5,1.9) {$B$};
%\node at (-.75,-1.9) {$C$};
%\node at (1.,-1.9) {$D$};
%\node at (-2.2,1.2) {in};
%\node at (-2.2,-1.2) {out};
%\end{tikzpicture}
\caption{To define tripartite information as a measure of scrambling the unitary operator $\hat U(t)$ is viewed as a state in doubled Hilbert space with \emph{in} and \emph{out} degrees of freedom.}
\label{fig:u_partition}
\end{figure}

In the following we study tripartite information as a measure of scrambling as introduced in Ref. \cite{Hosur2016}.
For simplicity we consider systems consisting of two-dimensional local Hilbert spaces and a corresponding basis $\{\ket{i}\}$.
To define tripartite information the time evolution operator acting on a system consisting of $N$ lattice sites,
\begin{align}
	\hat U(t)=\sum_{i,j}u_{ij}(t)|i\rangle\langle j|
\end{align}
which would commonly be interpreted as a tensor with $N$ input and $N$ output legs as
depicted in Fig.\ \ref{fig:u_partition},
is thought of as a state in doubled Hilbert space,
\begin{align}
	\ket{U(t)}&=\frac{1}{2^{n/2}}\sum_{i,j}u_{ij}(t)\ket{j}_\text{in}\otimes\ket{i}_\text{out}
	\nonumber\\
	&=\frac{1}{2^{n/2}}\sum_{j}\ket{j}_\text{in}\otimes\hat U(t)\ket{j}_\text{out}\ .
	\label{eq:unitary_channel}
\end{align}
In this language the reduced density matrix of the input subsystem, $\hat\rho_\text{in}=\text{tr}_\text{out}(\ket{U(t)}\bra{U(t)})$, corresponds to a uniform ensemble of states of the physical system, whereas the reduced density matrix of the output subsystem, $\hat\rho_\text{out}=\text{tr}_\text{in}(\ket{U(t)}\bra{U(t)})$, corresponds to the time-evolved initial density matrix, $\hat\rho_\text{out}=\hat U(t)\hat\rho_\text{in}\hat U(t)^\dagger$.

In this view one can consider more general input ensembles $\hat\rho_\text{in}=\sum_j p_j\ket{\psi_j}\bra{\psi_j}$ given by probabilities $p_j$ and a set of orthonormal states $\{\ket{\psi_j}\}$. The corresponding state $|\Psi(t)\rangle=\sum_j\sqrt{p_j}\ket{\psi_j}_\text{in}\otimes\hat U(t)\ket{\psi_j}_\text{out}$
contains all information about the time evolution, in this case with a possible weighting of the input ensemble. The following discussion is, however, restricted to the uniform ensemble corresponding to infinite temperature.

In the doubled system it is possible to define mutual information of subsystems on the input and on the output  
side. Considering bipartitions of the input and the output subsystem into parts $A,B,C$, and $D$ as
depicted in Fig.\ \ref{fig:u_partition} the mutual information of, e.g., $A$ and $C$ is defined as $I(A:C)=S_A+S_C-S_{AC}$
where $S_A=-\text{tr}\big(\hat\rho_A\log\hat\rho_A\big)$
with the reduced density matrix of subsystem $A$,
$\hat\rho_A=\text{tr}_{BCD}\big(|\Psi(t)\rangle\langle\Psi(t)|\big)$.
On this basis the tripartite information
\begin{align}
	I_3(A:C:D)=I(A:C)+I(A:D)-I(A:CD)\label{eq:i3}
\end{align}
quantifies how much information about $A$ is after time evolution hidden non-locally in $CD$ and
cannot be detected by local measurements just on $C$ or $D$. 
If a system scrambles initially local
information the tripartite information will assume a negative value with large magnitude. Therefore,
in contrast to OTOCs tripartite information allows to diagnose scrambling based only on properties
of the time evolution operator avoiding ambiguities that can occur due to the choice of observables.

%Note that tripartite information \eqref{eq:i3} defined on the doubled Hilbert space 
%is formally identical with the topological entanglement entropy \citep{KitaevPreskill2006}.
%Until now it is, however, unclear whether this resemblance has further implications.

When considering the infinite temperature ensemble, the density matrix $\hat\rho_{AB}(t)$ is proportional to the identity at all times; therefore, $I(A:CD)=2a$ with $a$ the size of subsystem $A$ is constant and, moreover, an upper bound for $-I_3(A:C:D)$ due to the positivity of mutual information. Making use of the time-independence of $\hat\rho_{AB}(t)$, Eq.\ \eqref{eq:i3} simplifies to $I_3(A:C:D)=N-S_{AC}-S_{AD}$.
Furthermore, writing the initial state $\ket{U(t=0)}$ at infinite temperature as a product of maximally entangled pairs in the \emph{in} and the \emph{out} part of the system, it is straightforward to show that initially $S_{AC}+S_{AD}=N$. Hence, under scrambling dynamics, the \emph{negative} tripartite information $-I_3$ will rise from zero to a large value, see also Ref.\ \cite{Hosur2016,Ding2016}. %\textcolor{red}{MS: is this the correct reference? SK: I would suggest to also include this reference from Hosur et al.}. 
Next, we discuss how the evolution with Haar random unitaries can be used as quantitative reference for values of tripartite information.

\section{Reference for scrambling.}
As a reference for scrambling we consider evolution with Haar random unitary operators
% , because these operators are expected to fully mix the information about the initial state; in this sense random unitaries correspond to maximally chaotic dynamics
\cite{Hosur2016,Zhou2017}.
For our analysis we will compute the value of the tripartite information attained under Haar scrambling numerically by considering the tripartite information in states $\ket{U_H}$ defined as in Eq. \eqref{eq:unitary_channel} with Haar random unitaries $\hat U_H$. We will consider a system a scrambler if the corresponding tripartite information is close to the average tripartite information obtained in the Haar ensemble of unitaries irrespective of the partitioning of the Hilbert space and the choice of subsystems $A,B,C$, and $D$. 
% A single choice 
The existence of a basis and a partitioning into subsystems where $-I_3$ remains well below the Haar scrambled value implies that information is not fully scrambled.
% On the other hand, the approach of $-I_3$ to the Haar scrambled value observed for a single partitioning does not allow to conclude that the system is a scrambler.
Note that $-I_3$ can exceed the Haar value as already pointed out in \cite{Hosur2016,Ding2016}.

To evaluate the Haar scrambled value of the tripartite information, we take a sample $\hat U_H$ from the unitary matrices\cite{Mezzadri2007}, calculate the tripartite information of the resulting state $\ket{U_H}$ (cf. Eq.\ \ref{eq:unitary_channel}) and then average over many samples. 

However, because of symmetries our model Hamiltonians, and hence also the corresponding time-evolution operators, are block matrices. This observation requires us to similarly use \emph{block random unitaries} for the Haar scrambled value.
To achieve this we take the block structure from the physical time evolution operator and fill each block with a random unitary matrix from the Haar measure.

Taking these symmetries into account we arrive at the Haar scrambled values used as references in this work. Interestingly, in this case it is sufficient to only consider the $N+1$ blocks of the particle number conservation, since further symmetries (momentum conservation, parity conservation) did not suppress the reference value any further.

\section{Model Hamiltonians and numerical method.}
For the purpose of this study we consider the following model Hamiltonian of spinless fermions in a one-dimensional lattice with periodic boundary conditions:
\begin{align}
\hat H(t_h,\lambda,V)&=-\frac{t_h}{2}\sum_{l=1}^{N}\big(\hat c_l^\dagger \hat c_{l+1}+\lambda \hat c_l^\dagger \hat c_{l+2}+h.c.\big)\nonumber\\
&\quad
+V \sum_{l=1}^{N}\hat c_l^\dagger \hat c_l\hat c_{l+1}^\dagger\hat c_{l+1} \, ,
\label{eq:ham}
\end{align}
 with $c_l^{(\dagger)}$ the usual fermionic annihilation (creation) operators on lattice site $l$.
%\textcolor{red}{SK: Where in the text do we consider which boundary conditions? E.g. in momentum space this remains unclear.}
%In the following we will consider both open and periodic boundary conditions.
By adjustment of the different parameters the system can be tuned between a quadratic Hamiltonian, a Bethe integrable system, and a generic non-integrable Hamiltonian. For $V=0$ the Hamiltonian is quadratic, irrespective of the value of $\lambda$. Any $V\neq0$ will add interactions to these fermions, but $\hat H(J,\lambda=0,V)$ is still integrable in the sense that it is solvable by Bethe ansatz \cite{Bethe1931}. Integrability is broken if $\lambda$ and $V$ are both nonzero. In the following we will fix $\lambda=0.5$ in order to contrast the behavior found in the non-integrable model against the quadratic model. 
While the focus will be on the quadratic and the non-integrable cases,
 results for the Bethe integrable system $\hat H(J,\lambda=0,V\neq0)$ are included in Section \ref{sec:Bethe}.

As an alternative partitioning of the Hilbert space we consider the system in momentum space, where
\begin{align}
\hat H(t_h,\lambda,V)&=
-t_h\sum_{k}\big(\cos(k)+\lambda\cos(2k)\big)\hat c_{k}^\dagger\hat c_{k}
\nonumber\\&\quad
-\frac{V}{N}
\sum_{k,k',q}\cos(q)
\hat c_{k+q}^\dagger\hat c_{k'-q}^\dagger\hat c_k\hat c_{k'}
\label{eq:ham_mom}
\end{align}
with $\hat c_{k_n}=N^{-1/2}\sum_le^{\im k_nl}\hat c_l$ and $k=2n\pi/N, n=0,\ldots, N-1$.

To obtain the dynamics of tripartite information \eqref{eq:i3} we compute numerically the exact time evolution in the full Hilbert space of systems with up to $N=12$ physical lattice sites, which means $2N=24$ sites in the doubled Hilbert space introduced in Eq.\ \eqref{eq:unitary_channel}. In order to reach these system sizes we avoid dealing with the full $2^{2N}$-dimensional state by directly computing the reduced density matrices of interest (see Appendix \ref{sec:numerics} for details). We checked our approach against results for the dynamics obtained with a method based on matrix product states \cite{Schollwoeck2011,tDMRGreviewPaeckel} and found good agreement. However, using that approach the strongly entangled initial state renders the simulation of long-time dynamics as required for the purpose of this work prohibitively expensive if the system size exceeds $N=12$ sites.

In Section \ref{sec:trip_inf_dyn} we present results for the time evolution of tripartite information in real and momentum space, followed by a detailed analysis of the asymptotic values at late times and finite size effects in Section \ref{sec:fs}. The findings allow us to conclude that in the limit of infinite system size the delocalization of information under dynamics of the non-integrable model is compatible with Haar scrambling for all partitionings of Hilbert space under consideration. For the quadratic model, instead, information remains more localized at all times, especially in momentum space, where the dynamics of tripartite information is trivial. Results for the Bethe integrable system are presented in Section \ref{sec:Bethe}.

\section{Time evolution of tripartite information}\label{sec:trip_inf_dyn}
\subsection{Real space}
\begin{figure}[t]
\begin{center}
\includegraphics[width=.95\columnwidth]{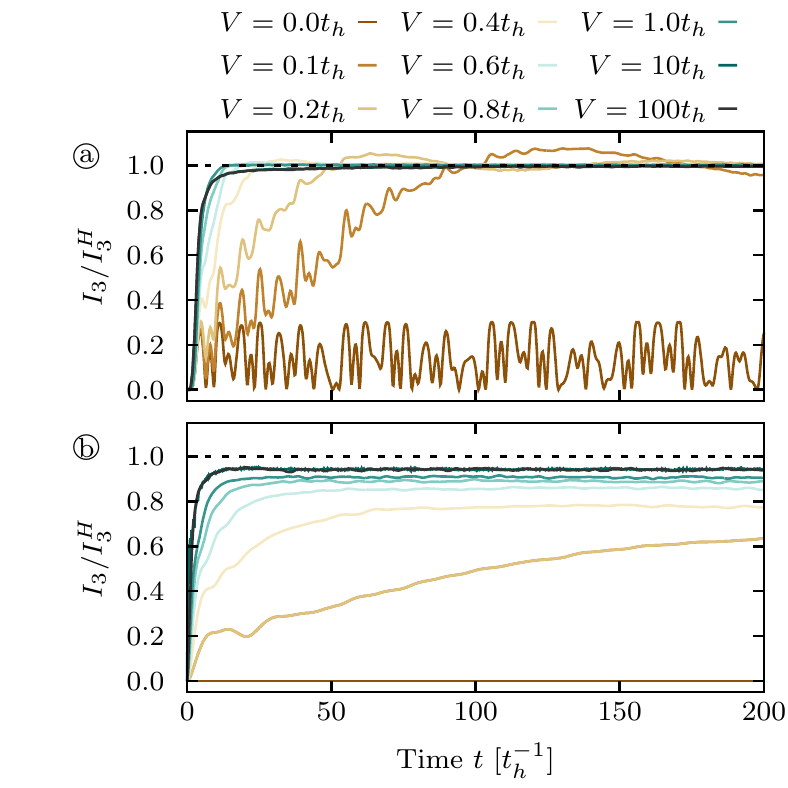}
\caption{Time evolution of the negative tripartite information in a chain of $N=12$ sites for $\lambda = 0.5$ and different values of the %integrability-breaking 
interaction $V$. (a) In real space, where $A$ and $D$ are single-site subsystems on diametrically opposing sides of the ring. (b) In momentum space, where $A$ is the mode $n=0$ and $D$ the mode $n=11$. The dashed lines indicate the Haar scrambled value of the tripartite information. 
%In both cases $\lambda=0.5$.
} 
\label{fig:scrambling}
\end{center}
\end{figure}

For the system with periodic boundary conditions we consider a partitioning of real space with single-site subsystems $A$ and $D$ located on diametrically opposing sides of the ring. In this setting the tripartite information shows characteristic differences depending on the Hamiltonian parameters, which is shown for a choice of parameters in Fig.\  \ref{fig:scrambling}a. 
In all cases the existence of a finite \emph{butterfly velocity} $v_B$ in real space is reflected in the fact that the tripartite information only deviates considerably from the initial value at time $t=l/v_B$, where $l$ is the distance between $A$ and $D$. This feature is not well resolved in Fig.\  \ref{fig:scrambling}, but will be further discussed in Section \ref{sec:bfv}.

Under time evolution with the quadratic Hamiltonian $H(1,0.5,0)$ the tripartite information shows a distinct signal for a short time at $t\approx l/v_B$, which subsequently decays, before revivals occur at later times. For $t>l/v_B$ the dynamics is characterized by strong oscillations. On average, however, the tripartite information remains well below the Haar scrambled value (dashed line).
By contrast, under dynamics of the interacting model the negative tripartite information rapidly raises to the Haar scrambled value at $t\approx l/v_B$ and does not deviate from that in the subsequent evolution.

\subsection{Momentum space}
In momentum space the many-body basis can be chosen as the set of Fock states characterized by momentum mode occupation numbers. Information is delocalized when it is distributed over the different modes $n$. Accordingly, when computing tripartite information the external legs of the time evolution operator in Fig.\ \ref{fig:u_partition} correspond to momentum mode indices. 
Note that the time evolution operator in momentum space has an additional block structure due to the conservation of total momentum, which needs to be taken into account in the corresponding Haar random unitary.

Fig.\ \ref{fig:scrambling}b displays the dynamics of the tripartite information in momentum space for different interaction strengths $V$. For the quadratic Hamiltonian with $V=0$ the tripartite information remains zero for all times. This is due to the fact that in this case the initial product structure is preserved in $\ket{U(t)}$; the state remains a product of maximally entangled pairs at all times, leaving tripartite information unchanged.

By contrast, the tripartite information under evolution with the interacting Hamiltonian quickly approaches a stationary value close to but below that obtained when evolving with a random unitary. The stationary value attained at late times in the presence of interactions is clearly distinct from the Haar scrambling value indicated by the dashed line. However, in Section \ref{sec:fs} we include a finite size analysis indicating that in the thermodynamic limit the asymptotic value is compatible with the Haar scrambling value.

We find that there is no butterfly velocity in momentum space as the tripartite information starts to deviate from the initial value immediately, irrespective of the choice of subsystems, because the Hamiltonian in momentum space \eqref{eq:ham_mom} has no notion of neighborhood. Instead, we find that the timescale for the increase of $-I_3$ is proportional to the interaction parameter $V$. 
%We attribute the stronger fluctuations of $I_3$ that occur for much longer times than in real space to the finite system size that is effectively reduced due to the conservation of momentum: the largest block in the Hamiltonian is much smaller than in the case without translational symmetry, which reduces the dimension of the subspaces that are scrambled in the course of the dynamics. We expect that these fluctuations vanish in the thermodynamic limit $N\to\infty$.

\subsection{Butterfly velocity and wave front broadening}\label{sec:bfv}
\begin{figure}[h]
\begin{center}

\vspace{-1cm}
\includegraphics[width=.95\columnwidth]{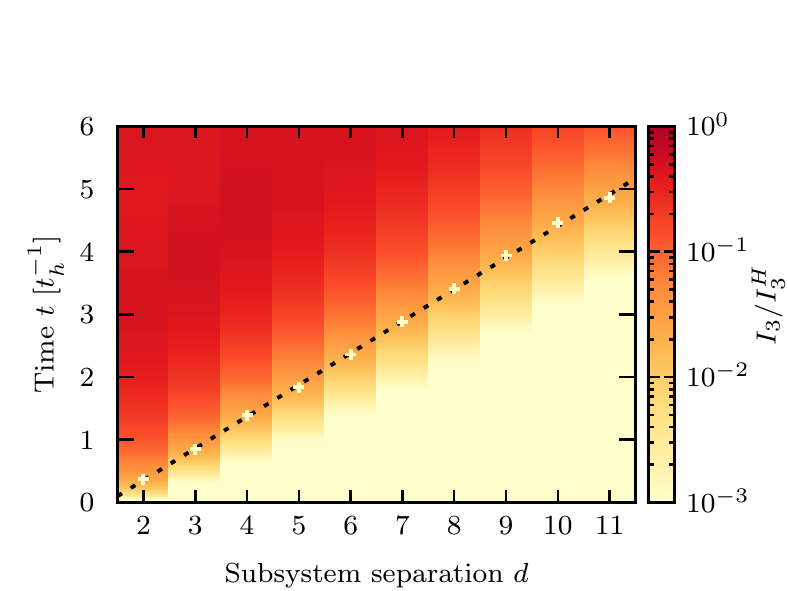}
\caption{Dynamics of tripartite information in the space-time plane. The plot shows the time evolution of the tripartite information with varying distance between the subsystems $A$ and $D$ for a system of $N=12$ lattice sites in the non-integrable regime, $\lambda=0.5$ and $V=0.5t_h$. The crosses mark the points at which $-I_3=0.031$ and the dashed line is a linear fit to these points yielding a butterfly velocity of $v_B \approx 1.97 t_h$.}
\label{fig:butterfly_velocity}
\end{center}
\end{figure}

In real space the tripartite information shows a clear signature of a \emph{light cone} as the separation between the subsystems $A$ and $D$ is varied. The corresponding characteristic velocity has been dubbed \emph{butterfly velocity}. In Fig.\ \ref{fig:butterfly_velocity} we show exemplarily the time evolution of the tripartite information with varying distance between the subsystems $A$ and $D$ for a system of $N=12$ lattice sites in the non-integrable regime, $\lambda=0.5$ and $V=0.5t_h$. The crosses mark the points at which the tripartite information grows beyond the threshold of $-I_3=0.031$ and a linear fit to these points yields a butterfly velocity of $v_B\approx1.97t_h$. We considered different values for the threshold and chose this particular one, because the deviation from linearity was minimal with this value.

\begin{figure}[b]
\begin{center}
\includegraphics[width=.95\columnwidth]{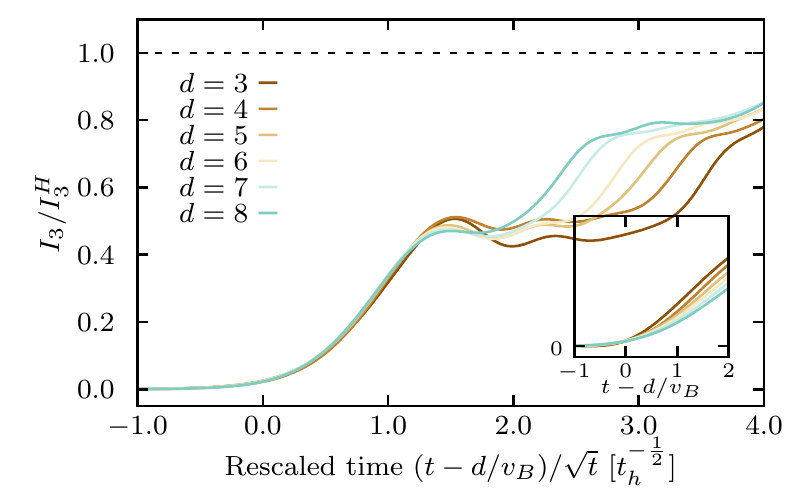}
\caption{Broadening of the wave front for the same parameters as in Fig.~\ref{fig:butterfly_velocity}. The data obtained for the dynamics of the tripartite information with different distances $d$ between subsystems coincide after shifting according to the butterfly velocity $v_B$ and a rescaling by $t^{-1/2}$ to account for \emph{diffusive broadening}. The inset shows the same data without a rescaling of the time axis for comparison.}
\label{fig:broadening}
\end{center}
\end{figure}

The dynamics of OTOCs exhibits a \emph{diffusive broadening of the wave front}, i.e., the time window between the first deviation of the OTOC from the initial value and the approach to the final value increases as the square root of the time \cite{vonKeyserlingk2017, Rakovszky2017, Khemani2018, Sahu2018, Knap18}. Our results for the tripartite information are compatible with an analogous behavior. In Fig.\ \ref{fig:broadening} we show the evolution of the tripartite information for different separations $d$ between the subsystems, where the time axis is rescaled as $\tau=(t-d/v_B)/\sqrt{t}$. After this rescaling the data for all distances coincide very well for $\tau\lesssim1/\sqrt{t_h}$. The agreement gets worse at later times, which is due to the finite  system size. Boundary effects that propagate into the bulk impede the collapse of the data. These effects impact subsystems close to the boundary earlier, which is the reason why only distances $3\leq d\leq8$ are shown. Note that only for the analysis of the butterfly effect the periodic boundary conditions have been replaced with an open boundary, which allows larger distances between the subsystems $A$ and $D$.

\section{Finite size effects and sensitivity to the breaking of integrability}\label{sec:fs}
The results presented in the previous section raise questions about the asymptotic values attained by the tripartite information at late times. In this section we include a careful analysis of the dependence of these late time values on the interaction parameter $V$ and the system size $N$.

\begin{figure}[b]
\begin{center}
\includegraphics[width=.95\columnwidth]{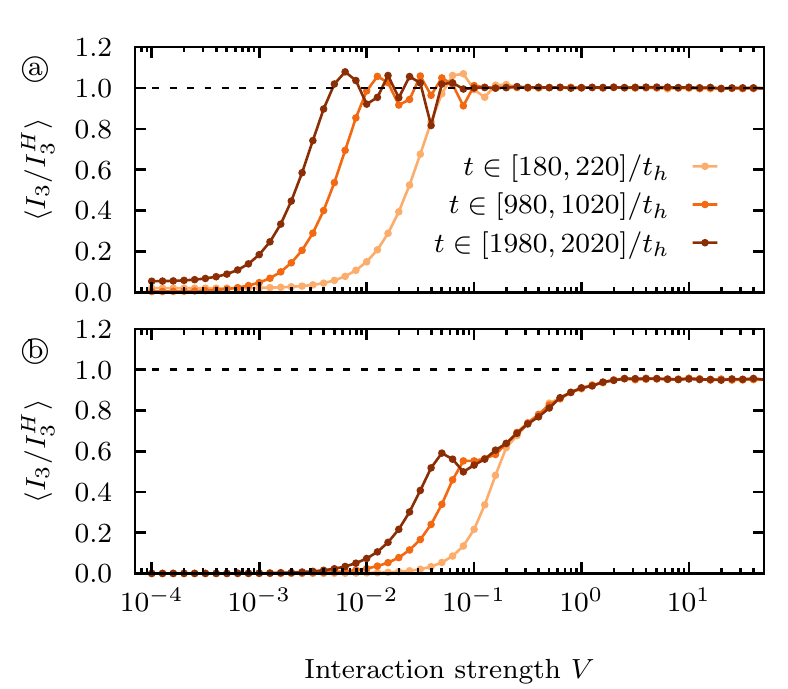} 
\caption{Averages of $-I_3$ taken over different intervals $[t_0,t_0+\Delta t]$ as a function of the 
% integrability breaking 
interaction parameter $V$ for $\lambda=0.5$. (a) In real space, where $A$ and $D$ are single-site subsystems on diametrically opposing sides of the ring. (b) In momentum space, where $A$ is the mode $n=0$ and $D$ the mode $n=11$. Here $N=12$ and the dashed line indicates the Haar scrambled value.
}
\label{fig:integrability_tuning}
\end{center}
\end{figure}
\subsection{Tripartite information at late times}
%The transition between the different characteristics at late times is very rapid as integrability is broken by tuning in the interaction. 
In Fig.\ \ref{fig:integrability_tuning}a we show averages of the tripartite information for the same real space partitioning as in Fig.\ \ref{fig:scrambling}a over certain intervals $[t_0,t_0+\Delta t]$ at late times $t_0$. These averages give an estimate of the stationary values attained in the long time limit. We find that when tuning to the interacting
model with $V>0$ the tripartite information quickly attains a new stationary value. 
Considering the small system sizes we study, this means that tripartite information is extraordinarily sensitive
to the presence of interactions. 
Notice that the transition to the Haar value occurs at smaller $V$ as $t_0$ is increased. 
% Therefore, 
We conjecture that the non-universal behavior for small $V$ is a finite system size effect and that for any non-vanishing $V$ the tripartite information will approach the Haar scrambled value for $t\to\infty$ in the thermodynamic limit.

Fig.\ \ref{fig:integrability_tuning}b shows the dependence of tripartite information in momentum space on the  interaction parameter $V$. The behavior is similar, but larger values of $V$ are needed for considerable deviations from zero. As already discussed in the previous section, the asymptotic value never reaches the Haar scrambling value. 
%\textcolor{red}{SK: I removed two sentences below that I found confusing. We can discuss this.}
% Moreover, in real space any deviation from the asymptotic value at large $V$ seems to originate in the averaging at finite time $t<\infty$. However, in momentum space there is a deviation from the value at large $V$ for $0.1t_h\lesssim V\lesssim t_h$. 
However, in the following section we present a systematic finite size analysis that is compatible with convergence to the Haar scrambling value in the thermodynamic limit for both $V\lesssim t_h$ and $V\gg t_h$.

\begin{figure}[t]
\begin{center}
\includegraphics[width=.95\columnwidth]{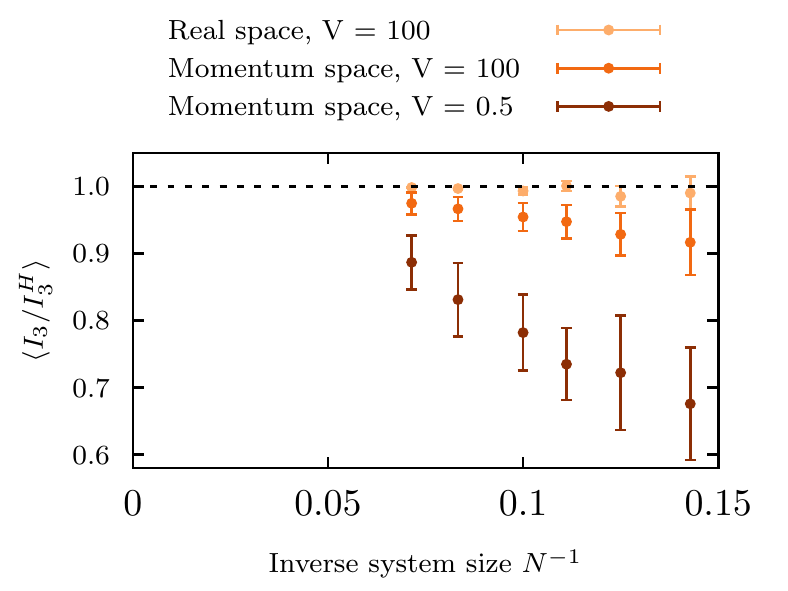} 
\caption{Averages of $-I_3$ at late times ($t > 2000$) for different inverse system sizes $N^{-1}$ for $\lambda = 0.5$. The $A$ and $D$ subsystems are of minimal size again, but were averaged over all possible choices of sites/momentum modes, which causes the larger errors in the momentum data. The dashed line indicates the Haar scrambled value.
}
\label{fig:finite_size_analysis}
\end{center}
\end{figure}

\subsection{Finite size analysis}
The results presented so far  show that information in momentum space
is not as effectively scrambled as in position space because the asymptotic values of the tripartite information in momentum space remains below the corresponding Haar scrambling value.

In Fig.\ \ref{fig:finite_size_analysis} we show asymptotic values of the tripartite information at late times that were estimated in the same way as in the previous section. The error bars reflect the fluctuations of the tripartite information on the time interval that is averaged over, as well as the dependence of the tripartite information on the choice of subsystems, which dominates in the momentum case. The data include two different values of the  interaction parameter $V$ for tripartite information in momentum space. In both cases we see that with increasing system size the late time values systematically approach the Haar scrambling value. Hence, given the system sizes that are accessible with our computational resources, we can conclude that the evolution of tripartite information in momentum space is compatible with Haar scrambling in the thermodynamic limit.

By contrast, the late time value found in real space is already for small finite systems close to the Haar scrambling value. We attribute this difference between real space and momentum space to the presence of an additional block structure of the time evolution operator in the momentum basis, namely total momentum blocks, which effectively reduces the degree of scrambling achievable in a finite system.

\begin{figure}[t]
\begin{center}
\includegraphics[width=.95\columnwidth]{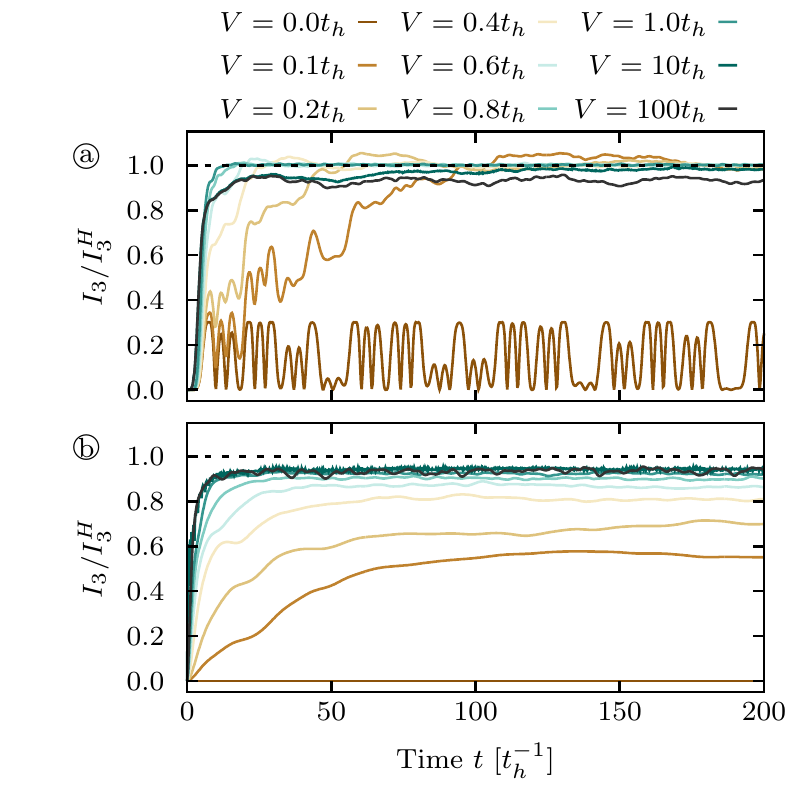}
\caption{Time evolution of the negative tripartite information in a chain of $N=12$ sites in the integrable regime with $\lambda=0$ for different values of the interaction $V$. (a) In real space, where $A$ and $D$ are single-site subsystems on diametrically opposing sides of the ring. (b) In momentum space, where $A$ is the mode $n=0$ and $D$ the mode $n=11$. The dashed lines indicate the Haar scrambled value of the tripartite information.}
\label{fig:I3Bethe}
\end{center}
\end{figure}
\section{Tripartite information in the Bethe integrable system}\label{sec:Bethe}

In Fig.\ \ref{fig:I3Bethe} we show the dynamics of the tripartite information in real space and momentum space for different interaction strengths $V$ with $\lambda=0$. Due to the absence of next-nearest neighbor hopping the systems are integrable for all values of $V$.

In momentum space, Fig.\ \ref{fig:I3Bethe}b, we find that the dynamics is very similar to the dynamics obtained with $\lambda=0.5$, shown in Fig.\ 2b. The reason is that in momentum space the deviation of $I_3$ from zero is due to the scattering term in the Hamiltonian, which is unaffected by the range of hopping (cf.\ Eq.\ \ref{eq:ham_mom}).

However, in comparison to the result for the non-integrable system (Fig.\ 2a), the tripartite information in real space shown in Fig.\ \ref{fig:I3Bethe}a deviates much more strongly from the Haar scrambled value also at late times. 

It is possible that finite size effects play a more important role in the Bethe integrable model and that by investigating larger systems one would find behavior closer to the non-integrable model.
% could conclude that in this case information is also scrambled in real space. However, also in this case scrambling with respect to partitionings in both real and momentum space is necessary but not sufficient to characterize the system as a scrambler. Instead, it might be possible to find partitionings of the Hilbert space, where information remains local under dynamics. In particular, 
However, notice that for the integrable model even in the thermodynamic limit we expect that information is not scrambled with respect to partitionings constructed from the quasiparticle basis, because in this basis scattering only leads to the permutation of rapidities, which is insufficient for the scrambling of information. The corresponding analysis is, however, beyond the scope of this work.

\section{Discussion}
%\textcolor{red}{MS: Here we need to add something about the significance of the thermodynamic limit. SK: I partly rewrote the paragraphs below.}
In this work we emphasize the importance of studying scrambling with respect to different partitionings of Hilbert space,
% Based on our results we suggest to introduce the notion of \emph{scrambling with respect to a partitioning of Hilbert space}, 
$\mathcal H=\mathcal H_A\otimes\mathcal H_B=\mathcal H_C\otimes\mathcal H_D$. A system should only be considered a scrambler if it scrambles information with respect to any physically relevant partitioning of Hilbert space. A physically relevant partitioning is one where experimentally accessible observables can be constructed which act exclusively on the individual factors of the partitioned space; these are the meaningful partitionings, because any information that is localized in the corresponding subsystems can in practice only be accessed via such observables.
%Our results show that in order to decide whether information is scrambled under the dynamics of a given system it is essential to investigate how information spreads with regard to different partitionings of the Hilbert space. We propose that a system can only be considered a scrambler if it distributes information globally irrespective of the choice of a physically relevant basis and the spatial segmentation. A relevant basis is one that can be constructed as a product of few-body eigenspaces; this excludes for example the energy eigenbasis of the Hamiltonian, with regard to which no system is a scrambler.
Clearly, for any numerical study this notion of scrambling only allows for falsification.
The observation of scrambling with respect to a specific choice of Hilbert space partitionings is necessary, but not sufficient for genuine scrambling.

The numerical results presented in this work show that non-interacting fermions in one dimension do not scramble; in particular, in momentum space information is not distributed at all. By contrast, the behavior of $I_3$ obtained for interacting systems is compatible with scrambling.  
%them being denoted scramblers. 
% and for all considered partitionings in real and momentum space information is mixed to the same extent as under evolution with a corresponding Haar random unitary. We expect that the same holds for any other physically relevant partitioning.

Our data also indicates that in the thermodynamic limit generic (meaning non-integrable) interacting systems at long times scramble as effectively as Haar random unitaries. For interacting Bethe ansatz integrable models we are unable to address this question due to stronger finite size effects. 

The results presented in this work show that tripartite information, which goes beyond OTOCs in that it directly quantifies the distribution of information, is an insightful measure for scrambling; e.g., a sharp distinction of system characteristics was revealed within the dynamics of $I_3$. As such, tripartite information should be further explored in future research to enhance the understanding of scrambling, including, e.g., the role of temperature.

\begin{acknowledgments}
The authors acknowledge helpful discussions with T.\ K\"ohler and N.\ Abeling. This work was financially supported through SFB/CRC 1073 (project  B03) and by Research Unit FOR 1807 (project P7) of the Deutsche Forschungsgemeinschaft (DFG).  
M.S.\ acknowledges support by the Studienstiftung des Deutschen Volkes and through the Leopoldina Fellowship Programme of the German National Academy of Sciences Leopoldina (Grant No. LPDS 2018-07) with additional support from the Simons Foundation.
\end{acknowledgments}

%\bibliographystyle{plain}
%\bibliography{references}

\appendix
\section{Numerical approach}\label{sec:numerics}
For the analysis of the tripartite information at infinite temperature the entanglement entropies $S_{AC}$ and $S_{AD}$ of the state $\ket{U(t)}$ in the doubled Hilbert space are needed. A straightforward way to obtain these would be to compute the full time-evolved state $\ket{U(t)}$, form the corresponding density matrix $\hat\rho(t)=\ket{U(t)}\bra{U(t)}$ and trace out the respective complements to obtain the reduced density matrices and from these the entropies. However, with this approach compute resources restrict the feasible sizes of the physical system to $N\lesssim7$, i.e., a doubled system with $2N\lesssim14$ sites. In order to obtain the data for $N=12$ presented in the main text we chose an alternative approach.

In our approach we individually compute the contributions to the reduced density matrix, for which it is sufficient to evolve states in the physical system and not the doubled system. The time-evolved state in the doubled system is
\begin{align}
    \ket{U(t)}=\sum_i \ket{i}_{AB}\otimes\hat U(t)\ket{i}_{CD}\ .
\end{align}
It is then convenient to think of the corresponding density matrix $\hat\rho(t)$ as a matrix of dimension $2^N\times2^N$, where every entry $\rho_{ij}$ is the corresponding matrix $\ket{i(t)}\bra{j(t)}$ obtained from the time-evolved basis states $\ket{i(t)}=\hat U(t)\ket{i}$.
Using this form of $\hat \rho(t)$ the contributions to the reduced density matrices of interest are easily determined and they can be computed exactly based on the time-evolved basis states of the physical system, $\ket{i(t)}$, without ever dealing with the full density matrix $\ket{U(t)}\bra{U(t)}$.

We also pursued an approach based on a matrix product state (MPS) representation of the infinite temperature state with subsequent time evolution. Within this real space Ansatz class there is a direct access to the entanglement spectrum for any single cut bipartition of the physical system. To be able to also treat embedded subsystems we developed a permutation scheme based on exact matrix product operator representations of permutation operators. 
With this method we were able to confirm the exact calculations with lattice sizes of $N=10$ and $N=12$ where we kept a maximum number of $\chi=1000$ states.
However, extending the simulation to larger systems turns out to be very challenging due to the fact that the initial state of the time evolution has a volume law of the entanglement entropy. Even though this volume law can be hidden in the particular choice of the initial state rendering the time evolution tractable, the calculation of the required permutations yields subsystems in which the scaling of the entropy with the volume of the subsystem reenters the calculations.
In detail we calculated for $N=14$ the time evolution of $-I_{3}$ with maximal number of kept states $\chi = 1000, 1500, 2000, 2500$ but where not able to obtain a well-converged result. 
We want to point out that with different initial states, e.g. finite temperature states, these calculations may be doable and the benefits of the MPS representation can be exploited.

%
% I_3 Haar without symmetries:
% -1.4073
% I_3 Haar with particle number conservation:
% -1.340
%

%\begin{thebibliography}{0}
%\bibliographystyle{plain}
\bibliography{references}
%\end{thebibliography}

\end{document}